# Gap solitons on a ring

Yaroslav V. Kartashov,[1] Boris A. Malomed,[2] Victor A. Vysloukh,[3] and Lluis Torner[1]

[1]*ICFO-Institut de Ciencies Fotoniques, and Universitat Politecnica de Catalunya, Mediterranean Technology Park, 08860 Castelldefels (Barcelona), Spain*

[2]*Department of Physical Electronics, School of Electrical Engineering, Faculty of Engineering, Tel Aviv University, Tel Aviv, 69978, Israel*

[3]*Departamento de Fisica y Matematicas, Universidad de las Americas – Puebla, Santa Catarina Martir, 72820, Puebla, Mexico*

We introduce a new species of gap solitons (GSs) supported by an azimuthally modulated guiding ring in defocusing cubic media. The periodicity in the azimuthal direction strongly modifies properties and existence domains of GSs. In addition to the fundamental solitons, we report even and twisted bound states. The former type is found to be stable, while the twisted states are always unstable in defocusing media.

Since their prediction [1,2] and observation [3] in fiber Bragg gratings, gap solitons (GSs) have been found in a variety of physical media, including waveguides with the refractive index periodically modulated in the transverse plane. Such spatial GSs, supported by the balance of the nonlinearity, diffraction, and Bragg reflection from the periodic structure, were observed in optically-induced lattices [4-8] and waveguide arrays [9-11]. Gap soliton trains [7,11], "embedded" (in-band) GSs [8], and extended gap states supported by layered materials [12] have been reported too. GSs were predicted [13-15] and created [16] as localized matter waves in Bose-Einstein condensates loaded into optical lattices.

Since GSs may extend over many periods of the underlying periodic structure, a particularly interesting situation arises when the guiding structure is finite, being subject to relevant boundary conditions (b.c.; in particular, *periodic* b.c. in the *ring-shape*d setting) at the edges. Modifications of the spectrum of such circular guiding structures and properties of the respective GSs have not been investigated in defocusing media (previously studied were multipole states in focusing azimuthally modulated lattices that belong to this class of structures [17], as well as GSs created by *radial* propagation of light in a multi-ring Bragg structure [18]).

In this Letter we introduce the properties of GSs in the most fundamental setting, based on a *single* azimuthally modulated guiding ring imprinted into a defocusing cubic medium. The respective periodic boundary conditions in the azimuthal direction result in an unusual structure of existence domains for extended GSs residing in such a circular grating.

In the setting addressed, light propagation along the axial direction ($\xi$) obeys the nonlinear Schrödinger equation for dimensionless field amplitude $q$:

$$i\frac{\partial q}{\partial \xi} = -\frac{1}{2}\left(\frac{\partial^2 q}{\partial \eta^2} + \frac{\partial^2 q}{\partial \zeta^2}\right) + q|q|^2 - pR(\eta,\zeta)q, \qquad (1)$$

with transverse coordinates ($\eta,\zeta$) normalized to the characteristic beam's width, while $\xi$ is normalized to the diffraction length. Further, $p$ is the depth of the refractive-index modulation, and $R = \exp[-(r-r_0)^2/d^2]\cos(n_\phi\phi)^2$ is the modulation profile, written in terms of polar coordinates $r$ and $\phi$. Here $r_0$ is the radius of the guiding ring, $d$ its width, and $2n_\phi$ the number of modulation periods in the azimuthal direction. Such structures can be fabricated in suitable defocusing materials using tightly focused pulses producing an irreversible modification of the refractive index in the focal region. In guiding rings written in $LiNbO_3$, the formation of solitons with width $\sim 5\,\mu m$ at $\lambda = 532\,nm$ may be expected at power levels $\sim 10\,\mu W$. Note that $p \sim 10$ corresponds to a refractive index change $\delta n \approx 0.0013$, and $\xi = 50$ is equivalent to a crystal length of about 30 mm. Notice that azimuthally *chirped* structures were proposed as ring accelerators for confined wavepackets in focusing media [19]. We fix the scales by choosing $r_0 = 5$.

Gap soliton solutions to Eq. (1) with axial propagation constant $b$ are looked for as $q(\eta,\zeta,\xi) = w(\eta,\zeta)\exp(ib\xi)$. The soliton family is characterized by the dependence on $b$ of total power $U = \int\int_{-\infty}^{\infty}|q|^2\,d\eta d\zeta$. For the linear stability analysis, we substitute perturbed solution $q = [w + (u+iv)\exp(\delta\xi)]\exp(ib\xi)$ into Eq. (1) and linearize it, which yields an eigenvalue problem for instability growth rate $\delta$ and perturbation components $u,v$:

$$\delta u = -\frac{1}{2}\left(\frac{\partial^2 v}{\partial \eta^2} + \frac{\partial^2 v}{\partial \zeta^2}\right) + bv + vw^2 - pRv,$$
$$\delta v = \frac{1}{2}\left(\frac{\partial^2 u}{\partial \eta^2} + \frac{\partial^2 u}{\partial \zeta^2}\right) - bu - 3uw^2 + pRu. \quad (2)$$

Typical profiles of fundamental GSs on the ring, belonging to the first finite bandgap of the linear spectrum, are shown in Fig. (1). The solitons are trapped around $r = r_0$ in the radial direction due to the total internal reflection, i.e., the presence of annular waveguide. In contrast, in the azimuthal direction the localization is accounted for by the Bragg reflection, hence the GSs feature characteristic oscillating tails.

The localization in the azimuthal direction depends on $b$ value, in terms of which there exist lower and upper cutoffs, $b_{\text{low}}$ and $b_{\text{upp}}$. The soliton's power monotonically decreases with the increase of $b$ [Fig. 2(a)]. For sufficiently large $p$ (deep azimuthal lattice), the soliton's amplitude increases at $b \to b_{\text{low}}$, and the tails become longer, until they touch each other [Fig. 1(a)], hence the solution ceases to be localized. One may identify the corresponding value of $b$, $b_{\text{low}}$, as a lower edge of the first finite bandgap, although, strictly speaking, a continuous bandgap spectrum does not exist in finite systems. Actually, "anti-dark" solitons exist in the present system at $b < b_{\text{low}}$, built as narrow peaks on top of an azimuthally modulated circular pedestal. Accordingly, the $U(b)$ dependence abruptly changes its slope at $b = b_{\text{low}}$ (the part with $b < b_{\text{low}}$ is not shown here, as the GSs belonging to that part tend to be unstable). In the middle of their existence interval GSs are well localized [Fig. 1(b)], but, at $b \to b_{\text{upp}}$, the field again expands across the entire ring, the peak amplitude decreases, and the GS transforms into a linear mode of the guiding ring (a counterpart of a Bloch state in infinite systems) [Fig. 1(c)].

The existence domain of GS on the ring expands with the increase of lattice's depth $p$, being drastically different from such domains for one- and two-dimensional GSs in infinite media [Fig. 2(b)]. As the medium is uniform at $r \to \infty$, spatially localized states can exist only with $b > 0$, in contrast to usual GSs, that can exist at $b < 0$. For $p$ sufficiently small, lower cutoff $b_{\text{low}}$ approaches zero. In this case, the GS with $b \to b_{\text{low}}$ does not expand across the entire ring; instead, it develops long tails penetrating into the uniform medium (in the radial direction), which is accompanied by an abrupt increase of the total power. Further decrease of $p$ results in disappearance of GSs, i.e., they exist only above a certain minimum value of $p$. The increase of the number of azimuthal modulation periods $2n_\phi$ first causes expansion of the existence domain, followed by its complete shrinkage; for the set of parameters used in Fig. 2(c), the existence region closes down at $n_\phi > 20$. The azimuthally modulated circular system may be considered as a quasi-one-dimensional one only under specific conditions, as ordinary one-dimensional lattices with arbitrarily small depth and period support gap solitons.

The modification in the slope of $U(b)$ at $b = b_{\text{low}}$ becomes less pronounced with the decrease of $n_\phi$: one can clearly distinguish two branches of the $U(b)$ dependence, corresponding to the azimuthally localized states and those extended across the ring, at $n_\phi > 10$, but not at smaller $n_\phi$. The same feature manifests itself through the modification of the $b_{\text{low}}(n_\phi)$ dependence around $n_\phi = 10$ [Fig. 2(c)]. Thus, $n_\phi = 10$ may be approximately identified as a border at which the system's behavior changes from that typical to circular arrays of several coupled waveguides into a regime emulating infinite periodic system. The width of the guiding ring, $d$, also strongly affects the GS existence domain [Fig. 2(d)]. Very narrow guiding rings can not support gap solitons, while both $b_{\text{low}}$ and $b_{\text{upp}}$ monotonically grow with $d$.

We also found in-phase [even, Fig. 3(a)] and out-of-phase [twisted, Fig. 4(a)] bound states of GSs residing in adjacent azimuthal channels. Their existence domains almost coincide with those for the fundamental solitons, with the total power also being a monotonically decreasing function of $b$ [Fig. 2(a)]. Naturally, curves $U_e(b)$ and $U_t(b)$ for the even and twisted states go, respectively, slightly above and below dependence $2U_f(b)$ for the fundamental solitons.

The stability analysis demonstrates that fundamental GSs on the ring are stable in the entire existence domain, except for a narrow region close to the lower cutoff (at high values of $U$), where the solitons suffer from a very weak oscillatory instability. Even bound states are also stable in a larger part of their existence domain, except for a narrow region close to the low-power cutoff, where an instability domain with a complex internal structure can be located [Fig. 2(f)]. Thus, for parameters of Fig. 2(f) the instability occurs at $5.66 \leq b \leq 6.22$, while the entire existence domain of the even bound states is $0.74 \leq b \leq 6.28$. An example of stable evolution of a perturbed even bound state is shown in Fig. 3. In contrast, twisted bound solitons are subject to a strong symmetry-breaking instability in their entire existence domain [Fig. 2(e)]. If perturbed, they transform into fundamental GSs, with emission of radiation (Fig. 4).

In conclusion, we have predicted that azimuthally modulated guiding rings imprinted into defocusing media support gap solitons whose properties are strongly affected by the ring width and depth, and by the number of periods of the azimuthal modulation.

# References with titles

# References without titles

# Figure captions

Figure 1.  The absolute value of the field in fundamental gap solitons on the ring for (a) $b = 0.75$, (b) $b = 3.50$, and (c) $b = 6.39$. In all cases, $p = 16$, $n_\phi = 10$, $d = 0.5$. The red circle designates the azimuthally modulated guiding ring.

Figure 2.  (a) The total power versus the propagation constant of fundamental gap solitons and even bound states for $p = 16$, $n_\phi = 10$, $d = 0.5$. Circles correspond to profiles shown in Fig. 1. Existence domains of the fundamental solitons: (b) in the $(p,b)$ plane, for $n_\phi = 10$, $d = 0.5$; (c) in the $(n_\phi,b)$ plane, for $p = 16$, $d = 0.5$; (d) in the $(d,b)$ plane, for $p = 16$, $n_\phi = 10$. The real part of the perturbation growth rate versus the propagation constant for (e) twisted and (f) even bound states, for $p = 16$, $n_\phi = 10$, $d = 0.5$.

Figure 3.  The absolute value of the field in a stable even bound state with $n_\phi = 10$, $d = 0.5$, $p = 16$, and $b = 5.6$, at $\xi = 0$ (a) and $\xi = 500$ (b). A broadband noise with variance $\sigma_{\text{noise}}^2 = 0.01$ was added to the input configuration.

Figure 4.  The same as in Fig. 3, but for an unstable twisted bound state with $b = 6.1$. Configuration (b) is displayed at $\xi = 42$.

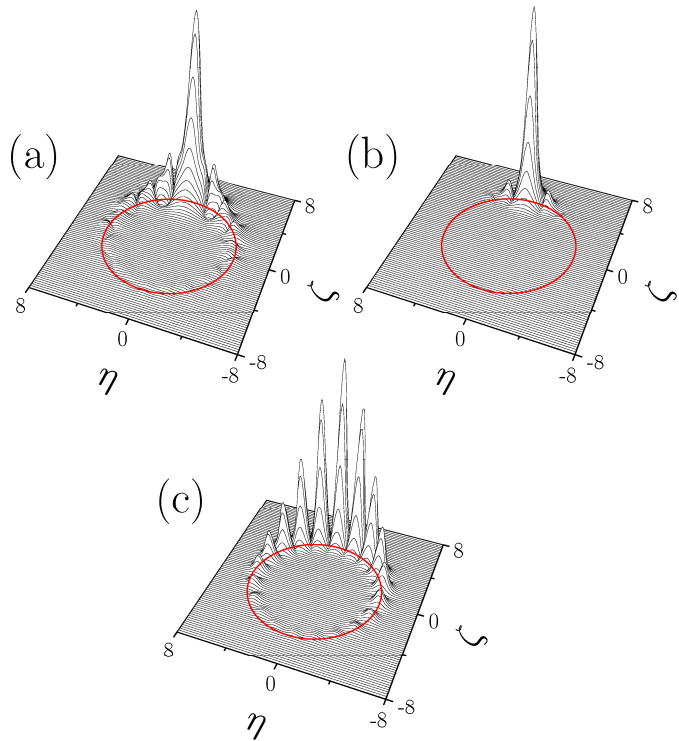

Figure 1. The absolute value of the field in fundamental gap solitons on the ring for (a) $b = 0.75$, (b) $b = 3.50$, and (c) $b = 6.39$. In all cases, $p = 16$, $n_\phi = 10$, $d = 0.5$. The red circle designates the azimuthally modulated guiding ring.

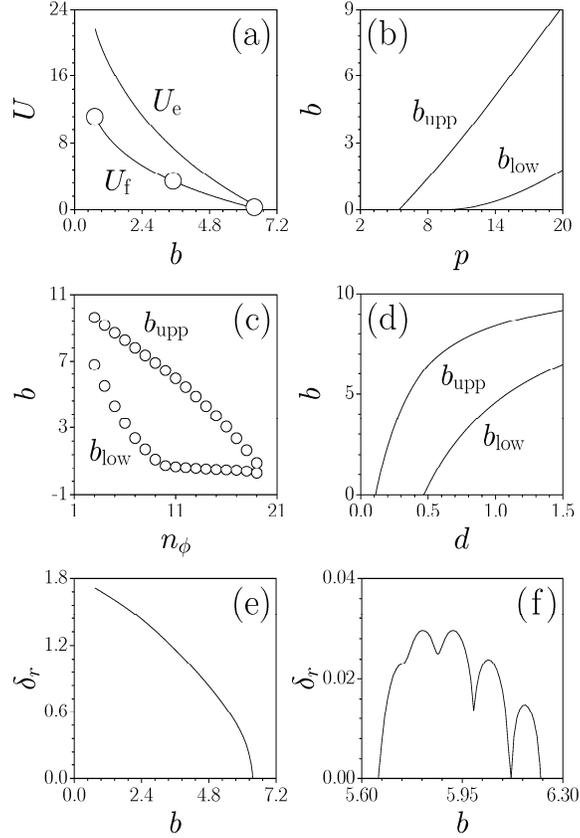

Figure 2. (a) The total power versus the propagation constant of fundamental solitons and even bound states for $p = 16$, $n_\phi = 10$, $d = 0.5$. Circles correspond to profiles shown in Fig. 1. Existence domains of the fundamental solitons: (b) in the $(p,b)$ plane, for $n_\phi = 10$, $d = 0.5$; (c) in the $(n_\phi,b)$ plane, for $p = 16$, $d = 0.5$; (d) in the $(d,b)$ plane, for $p = 16$, $n_\phi = 10$. The real part of the perturbation growth rate versus the propagation constant for (e) twisted and (f) even bound states, for $p = 16$, $n_\phi = 10$, $d = 0.5$.

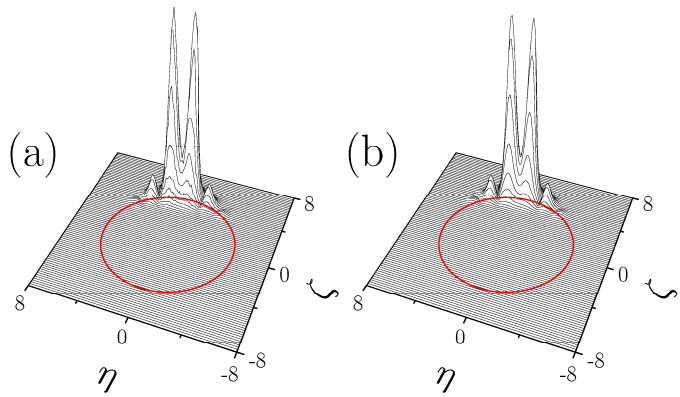

Figure 3. The absolute value of the field in a stable even bound state with $n_\phi = 10$, $d = 0.5$, $p = 16$, and $b = 5.6$, at $\xi = 0$ (a) and $\xi = 500$ (b). A broadband noise with variance $\sigma^2_{\text{noise}} = 0.01$ was added to the input configuration.

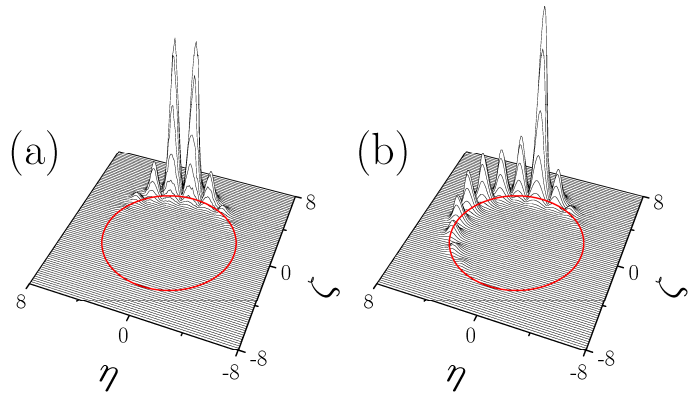

Figure 4. The same as in Fig. 3, but for an unstable twisted bound state with $b = 6.1$. Configuration (b) is displayed at $\xi = 42$.